\documentclass[aps,preprint,showpacs]{revtex4}
\usepackage{amssymb}
\usepackage{bm}
\usepackage{amsmath}
\usepackage{graphicx}

\begin{document}

\title{Three-dimensional spatiotemporal optical solitons in nonlocal
nonlinear media}
\author{D. Mihalache$^{1,2,4}$, D. Mazilu$^{1,2}$, F. Lederer$^{2}$, B. A.
Malomed$^{3}$, Y. V. Kartashov$^{4}$, L.-C. Crasovan$^{4,1}$, and
L. Torner$  ^{4}$}
\address{$^{1}$National Institute of Physics and Nuclear Engineering,
Institute of Atomic Physics, Department of Theoretical Physics, P.O. Box
MG-6, Bucharest, Romania\\
$^{2}$ Institute of Solid State Theory and Theoretical Optics,
Friedrich-Schiller Universit{\"a}t Jena, Max-Wien-Platz 1, D-077743 Jena,
Germany\\
$^{3}$Department of Interdisciplinary Studies, Faculty of Engineering, Tel
Aviv University, Tel Aviv 69978, Israel\\
$^{4}$ICFO - Institut de Ciencies Fotoniques, Mediterranean Technology Park,
08860 Castelldefels (Barcelona), Spain}

\begin{abstract}
We demonstrate the existence of stable three-dimensional spatiotemporal
solitons (STSs) in media with a nonlocal cubic nonlinearity. Fundamental
(nonspinning) STSs forming one-parameter families are stable if their
propagation constant exceeds a certain critical value, that is inversely
proportional to the range of nonlocality of nonlinear response. All spinning
three-dimensional STSs are found to be unstable.
\end{abstract}

\maketitle

Spatiotemporal solitons (STSs), also referred to as ``light
bullets" \cite  {Yaron}, have attracted a great deal of attention
in optics, see a recent review \cite{review}. These are
multidimensional pulses, which maintain their shape in the
longitudinal (temporal) and transverse (spatial) directions due to
the balance between the group-velocity dispersion, diffraction,
and nonlinear self-phase modulation. However, solitons in media
with the cubic self-focusing nonlinearity, obeying the nonlinear
Schr{\"{o}}  dinger (NLS) equation, are unstable in two and three
dimensions (2D and 3D), because of the occurrence of collapse in
the same model \cite{Berge}. Several possibilities to arrest the
collapse were considered, such as periodic alternation of focusing
and defocusing layers \cite{Isaac} and various generalizations of
this setting \cite{book}, and the use of media exhibiting
saturable \cite{saturable} or quadratic ($\chi ^{(2)}$) \cite
{quadratic} nonlinearities; \textit{tandem structures}, composed
of alternating linear and $\chi ^{(2)}$ layers, were also proposed
\cite{TLB}. The only successful experiment in this field was the
creation of quasi-(2+1)-dimensional STSs in bulk $\chi ^{(2)}$
samples \cite  {Frank,review}. Other theoretically developed
approaches use off-resonance two-level systems \cite{Igor} and
self-induced-transparency media \cite  {Miriam}.

Collapse does not occur either in $\chi ^{(3)}$ media whose
nonlinearity is nonlocal \cite{Turitsyn}, therefore they may also
give rise to stable solitons, see review \cite{nonlocal}. 2D
spatial solitons stabilized by the nonlocality were observed in
vapors \cite{vapor} and lead glasses featuring strong thermal
nonlinearity \cite{Moti}; in the latter case, elliptic and
vortex-ring solitons were reported. Optical 1D solitons supported
by a nonlocal $\chi ^{(3)}$ nonlinearity were also created in
liquid crystals \cite{liquid}. Further, self-focusing in
photorefractive media \cite  {Zozulya_1996}, periodic lattices
\cite{Yaroslav}, vortices \cite{vortex}, and spatial solitons in
soft matter \cite{Trillo} were considered in the
context of nonlocality. %In addition, it was
%shown that long-range cubic nonlinearity induced by long-range
%interactions between atoms carrying polarized magnetic momenta in
%effectively two-dimensional Bose-Einstein condensates also leads to
%the prediction of stable 2D solitons \cite {dipolarBEC}.

The objective of this work is to demonstrate that one-parameter
families of \emph{stable} (3+1)-dimensional STSs are possible in
media with nonlocal $  \chi ^{(3)}$ nonlinearity. For this
purpose, we consider a 3D model based on a general system of
coupled equations for the complex field amplitude $q$ and
nonlinear correction to the refractive index $n$ (see, e.g., \cite
{Krolikowski_2001}); in a normalized form, the equations are
\begin{eqnarray}
iq_{\xi }+(1/2)(q_{\eta \eta }+q_{\zeta \zeta }+Dq_{\tau \tau })+qn &=&0,
\notag \\
d(n_{\eta \eta }+n_{\zeta \zeta })-n+|q|^{2} &=&0.  \label{system}
\end{eqnarray}
Here, $\eta ,\zeta $ and $\xi $ are the transverse and longitudinal
coordinates, scaled, respectively, to the beam's width and its
diffraction length, $\tau $ is the reduced temporal variable, and
$D$ is the ratio of the diffraction and dispersion lengths. We
consider the case of \textit{\ anomalous} temporal dispersion,
$D>0$, and then set $D=1$ by means of an obvious scaling. Lastly,
$\sqrt{d}$ determines the correlation length $\Lambda
_{\mathrm{corr}  }$ of the nonlocal nonlinear response [note that by
setting $d=0$ one turns Eqs. (\ref{system}) into the ordinary 3D NLS
equation with self-focusing]. In fact, after setting $D=1$,
additional rescaling of Eqs. (\ref{system}) makes it possible to set
$d=1$, so as to cast the system into the parameter-free form.
Nevertheless, we prefer to keep $d$ as an explicit parameter, as it
directly controls the system's nonlocality degree.

The nonlocal nonlinearity in Eqs. (\ref{system}) is typical for
light propagation in liquid crystals and for thermal nonlinearity in
optical media \cite{nonlocal,liquid}. For the derivation of the
model equations (\ref{system}), the usual approximation of the
slowly varying amplitude is adopted, along with an assumption of
fast temporal relaxation of the refractive-index perturbations,
therefore the second equation in (\ref{system}) does not contain the
term $n_{\tau \tau }$. The latter assumption, which is essential for
the physical justification of the nonlocal model including the
temporal-dispersion term, implies that the relaxation time, $\tau
_{\mathrm{  rel}}\sim \Lambda _{\mathrm{corr}}/c$ ($c$ is the light
speed), must be of the order of the pulse duration $T$ of the STS to
be constructed. Whether such conditions can be met in available
optical media featuring nonlocal nonlinearities, such as liquid
crystals which tend to exhibit slower relaxation times, is a
question that remains a challenge.

Equations (\ref{system}) conserve the energy $E~=~\int \int \int
\left\vert q(\eta ,\zeta ,\tau )\right\vert ^{2}d\eta d\zeta d\tau$,
Hamiltonian $H$, the momentum $P_{\eta ,\zeta }$ in the transverse
plane, and angular momentum $M_{\xi}$ along the longitudinal
direction. Stationary solutions to Eqs. (\ref{system}) are looked
for as $q=w(r,\tau )\exp [i(b\xi +S\theta )]$, $n=n(r,\tau )$, where
$r$ and $\theta $ are the polar coordinates in the $\left( \eta
,\zeta \right) $ plane, $b$ is a real propagation constant, integer
$S$ is the vorticity (``spin"), and real functions $w$ and $n$ obey
the equations
\begin{eqnarray}
\left( w_{rr}+r^{-1}w_{r}+w_{\tau \tau }\right) -\left(
2b+r^{-2}S^{2}\right)w +2wn &=&0, \\
d\left( n_{rr}+r^{-1}n_{r}\right) -n+w^{2} &=&0
\end{eqnarray}
(for this solution, the angular momentum is proportional to the
energy, $  M_{\xi }=SE$).

We have numerically found families of localized solutions to these
equations, dealing with the corresponding two-point boundary-value
problem by dint of the standard band-matrix algorithm. Typically,
grids with $  241\times 240$ and $201\times 360$ points were used
for the computations of the fundamental ($S=0$) and spinning
solitons, respectively. In Fig. 1, we display the energy
characteristic, $E=E(b)$, for one-parameter families of the thus
constructed soliton solutions. In view of the above-mentioned
possibility to eliminate $d$ by means of rescaling, panel (a) in
Fig. 1 is, as a matter of fact, a blowup of a part of panel (b)
corresponding to $  0<b<2.4$. A noteworthy feature is that the 3D
solitons exist only above a finite energy threshold.

The stabilizing effect of the nonlocality for the fundamental
solitons is seen in Fig. 1: except for a narrow interval of small
wavenumbers, $0<b<b_{ \mathrm{cr}}$, the solitons are expected to
be stable, as they satisfy the \textit{Vakhitov-Kolokolov} (VK)
criterion, $dE/db>0$, which is a necessary (but, generally, not
sufficient) stability condition for the soliton family
\cite{VK,Berge} (note that the instability of the 3D solitons in
the local NLS equation precisely follows this criterion). The
bending of the curve $  b=b(E)$ for fundamental 3D solitons, from
the negative to positive slope, was also found in a different
model (a $\chi ^{(3)}$ medium with a Gaussian nonlocal kernel) by
means of a variational approximation \cite{Bang2002}. Below, it
will be demonstrated that the VK criterion is actually sufficient
for the stability of the fundamental solitons, but not for
spinning ones, which are always unstable.

For $d=1$, the critical wavenumber which separates the negative-
and positive-slope regions for the fundamental solitons in Fig. 1
is $b_{\mathrm{  \ cr}}^{(S=0)}=0.565$, and, in view of the
scaling invariance of the model, $
b_{\mathrm{cr}}^{(S=0)}(d)=0.565/d$. The energy of the $S=0$
soliton at the critical point is $E_{\mathrm{cr}}^{(S=0)}=42.60$
for $d=1$ (in the model with $d\neq 1$, $E_{\mathrm{cr}}$ scales
as $\sqrt{d}$, see Fig. 1). It also follows from here that
$E_{\mathrm{cr}}^{(S=0)}=\allowbreak 32.02\left( b_{
\mathrm{cr}}^{(S=0)}\right) ^{-1/2}$, which may be compared to the
scaling law for unstable fundamental solitons in the 3D local NLS
equation, $  E^{(S=0)}=\allowbreak C_{0}b^{-1/2}$, with
$C_{0}\approx 6.67$ \cite{Berge}.

Figures 2(a,b) display the shapes of typical stable and unstable fundamental
soliton for $d=10$, at a fixed value of the energy ($E=140$). It is seen
that high-amplitude solitons are stable, whereas low-amplitude ones are
unstable. We notice that both the low- and high-amplitude solitons feature
spatiotemporal ellipticity, being broader in space than in time, which can
be easily explained by a perturbative analysis of Eqs. (\ref{system}) for
small $d$. Typical shapes of unstable $S=1$ solitons are shown in Figs.
2(c,d).

Full stability of solitons was investigated using the equations for small
perturbations linearized around the stationary solution. Accordingly,
solutions including perturbations with an infinitesimal amplitude $\epsilon $
are looked for as
\begin{eqnarray}
q &=&e^{ib\xi +iS\theta }\left\{ w(r,\tau )+\epsilon \left[ f(r,\tau
)e^{\delta \xi +iJ\theta }+g^{\ast }(r,\tau )e^{\delta ^{\ast }\xi -iJ\theta
}\right] \right\} , \\
n &=&n(r,\tau )+\epsilon \left[ p(r,\tau )e^{\delta \xi +iJ\theta }+p^{\ast
}(r,\tau )e^{\delta ^{\ast }\xi -iJ\theta }\right] ,
\end{eqnarray}
where $J$ is an arbitrary integer azimuthal index of the
perturbation, $  \delta $ is the instability growth rate, the
asterisk stands for the complex conjugation, and the
eigenfunctions $f$, $g$ and $p$ obey the equations
\begin{eqnarray}
2i\delta f+f_{rr}+r^{-1}f_{r}+f_{\tau \tau }-\left[ 2b+(S+J)^{2}r^{-2}\right]
f+2\left( nf+wp\right) &=&0,  \notag \\
-2i\delta g+g_{rr}+r^{-1}g_{r}+g_{\tau \tau }-\left[ 2b+(S-J)^{2}r^{-2}
\right] g+2\left( ng+wp\right) &=&0,  \label{linear} \\
d(p_{rr}+r^{-1}p_{r})-(1+dJ^{2}r^{-2})p+w(f+g) &=&0.  \notag
\end{eqnarray}

The growth rate $\delta $ was found as an eigenvalue at which Eqs.
(\ref  {linear}) has a nonsingular localized solution. In Figs.
3(a), 3(b,c), and 3(d), we plot the instability gain,
$\mathrm{Re}(\delta )$, vs. the propagation constant $b$ of the
unperturbed solution, for the STSs with $S=0$ , $1$, and $2$,
respectively. The stable solitons are those for which $
\mathrm{Re} (\delta )=0$ for all (integer) values of $J$. Due to
the scaling invariance of Eqs. (\ref{system}), the curves in a
given panel, pertaining to different values of $d$, are obtained
from each other by the scaling transformation, therefore they
actually display essentially the same stability and instability
regions, but on different scales. Figure 3(a) shows
$\mathrm{Re}(\delta )$ for $J=0$, as $J\neq 0$ does not yield any
instability for the fundamental solitons; the stability region
revealed by this figure, $b>b_{\mathrm{cr}}$, is precisely the
same as predicted above by the VK criterion. For the spinning
solitons, the perturbations with $J>1$ destabilize a part of the
families, see, e.g., Fig. 3(c) for $S=1$ and $J=2$  , but entire
STS\ families with $S\geq 1$ are unstable against the
perturbations with $J=1$, see Fig. 3(b,d). The latter instability
mode implies a trend to splitting of the vortex soliton into a set
of two fundamental ones \cite{review}, which is corroborated by
direct simulations below.

Note that \emph{stable} 3D spinning optical solitons (with $S=1$)
were only found in media with \emph{competing} focusing and
defocusing nonlinearities, \textit{viz}., $\chi ^{(3)}:\chi
^{(5)}$ or $\chi ^{(2)}:\chi ^{(3)}$ \cite  {Mihalache_spinning}.
On the other hand, stable 2D spatial (rather than spatiotemporal)
vortex rings were experimentally observed in a medium featuring
the thermal nonlocal nonlinearity \cite{Moti}.

The predictions of the linear stability analysis were checked in direct
simulations of Eqs. (\ref{system}), which were run by means of a standard
Crank-Nicholson scheme. The nonlinear finite-difference equations were
solved using the Picard iteration method, and the resulting linear system
was handled with the help of the Gauss-Seidel iterative procedure. To
achieve good convergence, we needed typically, twelve Picard's and four
Gauss-Seidel iterations. The initial conditions for perturbed solitons were
taken as $q(\xi =0)=w(\eta ,\zeta ,\tau )(1+\epsilon \rho )$, and $n(\xi
=0)=n(\eta ,\zeta ,\tau )(1+\epsilon \rho )$, where $\epsilon $ is, as
above, a small perturbation amplitude, and $\rho $ was either a random
variable uniformly distributed in the interval $[-0.5,0.5]$, or simply $\rho
=1$ (uniform perturbation).

First, we have checked that all the fundamental STSs that were
predicted above to be stable (stable branches are shown with solid
lines in Fig. 1), are stable indeed against random perturbations;
Fig. 4 displays an example of self-healing of a stable soliton
with the initial perturbation amplitude $  \epsilon =0.1$. A small
uniform perturbation ($\rho =1$) applied to a stable soliton
excites its persistent oscillations, which suggests the existence
of a stable intrinsic mode in the soliton. On the other hand,
direct simulations show that those fundamental solitons that were
predicted to be unstable decay into radiation, if slightly
perturbed.
%%For the linearly unstable nonspinning solitons we have identified two
%%different instability scenarios: they either spread out under uniform
%%perturbations that reduce their norm (energy) or, or they collapse if the
%%soliton is perturbed by a relatively strong uniform perturbation with the
%%strength $\epsilon =0.05$, that increases their energy.
We also simulated self-trapping of a stable fundamental soliton from an
initial spatiotemporal pulse of an arbitrary form. An example is shown in
Fig. 5 for an isotropic Gaussian input, which generates an anisotropic
(elliptic) soliton. We also simulated the evolution of unstable spinning
solitons. Most typically, they follow the prediction of the linear stability
analysis and split into two stable fundamental solitons, see an example (for
$S=1$) in Fig. 6.

In conclusion, we have demonstrated that nonlocal cubic nonlinearity is
sufficient to stabilize 3D solitons, which suggests a new approach to making
of 3D spatiotemporal solitons in optics, which thus far evaded experimental
observation. The stability of the fundamental solitons was demonstrated
through the computation of the corresponding stability eigenvalues, and in
direct simulations. Their robustness and, hence, physical relevance was
demonstrated by self-trapping from arbitrary input pulses. On the other
hand, all the spinning 3D solitons in the nonlocal medium are unstable
against splitting into a set of stable fundamental solitons.

This work was partially supported by the Government of Spain through grant
BFM 2002-2861, by the Ramon-y-Cajal program, and by Deutsche
Forschungsgemeinschaft (DFG), Bonn. The work of B.A.M. was supported, in a
part, by the Israel Science Foundation through the grant No. 8006/03.

%\section*{References}

%\end{document}

\newpage

%\textbf{Figure Captions}

%\Figures

\begin{figure}
\caption{Energy $E$ vs. propagation constant $b$ for soliton
families with different vorticities, $S=0,1,$ and $ 2 $ (numbers
labeling the curves), at different values of the range of
nonlocality of nonlinear response $\protect\sqrt{d}$: (a) $d=1$,
(b) $d=10$. Full and dashed lines show stable and unstable
solitons.} \label{Fig.1}
\end{figure}

\begin{figure}
\caption{Cross-section shapes of typical $S=0$ and $S=1$ solitons
in the transverse ($r$) and temporal ($\protect\tau $) directions.
(a) and (b): $S=0$, $d=10$, $E=140$; full and dashed lines
correspond to stable and unstable solitons, with $b=0.15$ and
$b=0.021$, respectively. (c) and (d): $S=1$, $b=1$; solid and
dashed lines correspond to $d=1$ and $d=100$, respectively (due to
the scaling invariance, the latter is tantamount to $d=1$ and
$b=100$).} \label{Fig.2}
\end{figure}

\begin{figure}
\caption{The real part of the perturbation growth rate vs. the
propagation constant of the unperturbed soliton. (a) $S=0$, $J=0$,
for $d=1$ and $d=10$, (b) $S=1$, $J=1$, (c) $S=1$ , $J=2$, (d)
$S=2$, $d=20$. In (b) and (c), values of $d$, and in (d), values
of $J$ are indicated near the curves.} \label{Fig.3}
\end{figure}

\begin{figure}
\caption{Isosurface plots illustrating the stability of a
fundamental soliton corresponding to $d=10$, $b=1$, and $ E=178$.
(a) and (c): the initially perturbed soliton, at $\protect\xi =0$;
(b) and (d): the self-cleaned one at $\protect\xi =360$. Here and
in Figs. 5 and 6, the upper and lower rows show $|q|^2$ and $n$,
respectively.} \label{Fig.4}
\end{figure}

\begin{figure}
\caption{Self-trapping of a fundamental soliton, for $d=1$: (a)
and (c) -- the initial Gaussian pulse with energy $ E_{0}=54$; (b)
and (d) -- the soliton at $\protect\xi =60$.} \label{Fig.5}
\end{figure}

\begin{figure}
\caption{Splitting of an unstable $S=1$ soliton with $d=100$ and
$b=0.05$. (a) and (d) $\protect\xi =0$, (b) and (e) $\protect\xi
=1400$, and (c) and (f) $\protect\xi =1600$.} \label{Fig.6}
\end{figure}

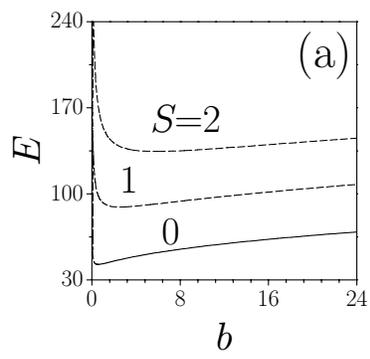
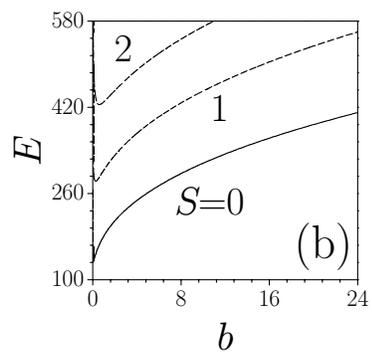
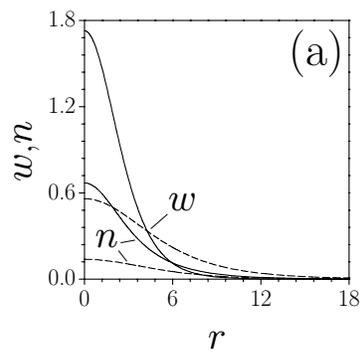
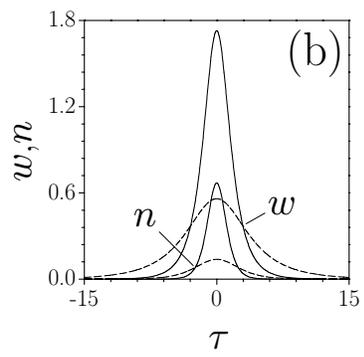
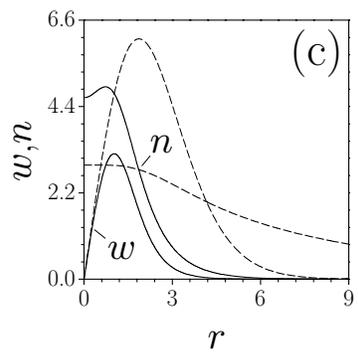
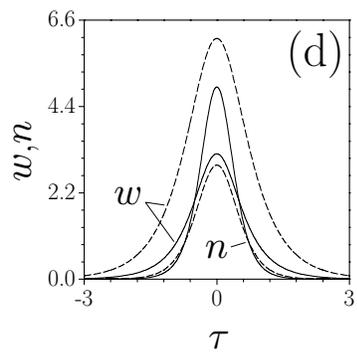
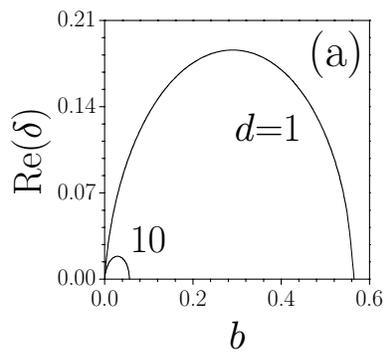
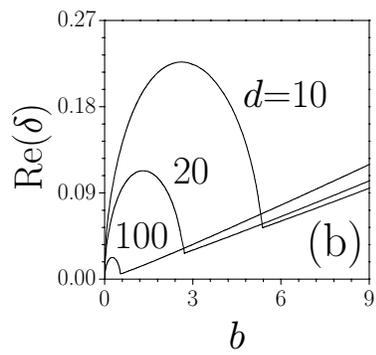
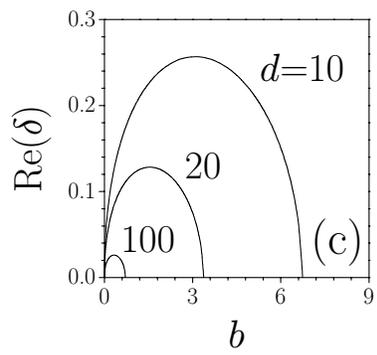
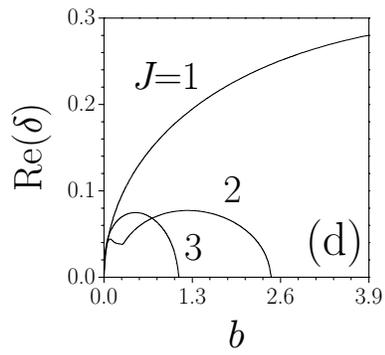
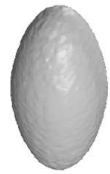
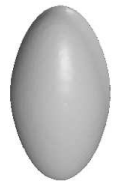
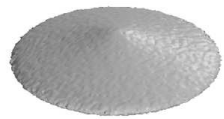
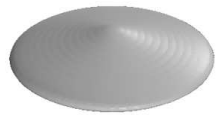
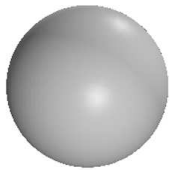
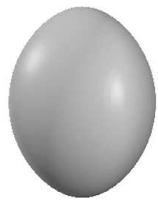
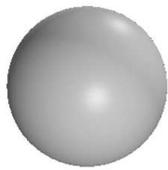
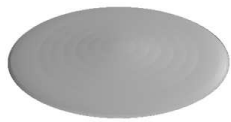
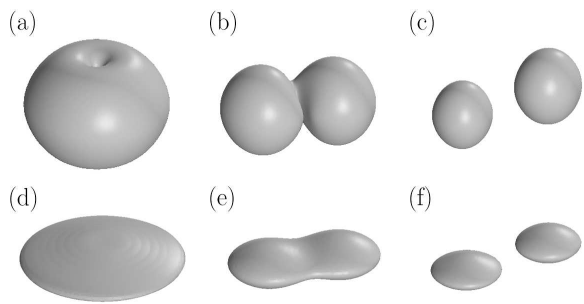
\end{document}